\documentclass{jnmp}
\begin{document}

\renewcommand{\evenhead}{G~Gaeta,~D~Levi,~R~Mancinelli}
\renewcommand{\oddhead}{Asymptotic symmetries of difference equations}

\def\a{\alpha}
\def\b{\beta}
\def\g{\gamma}
\def\ga{\gamma}
\def\de{\delta}   %% NON ridefinire come \d !!!!
\def\eps{\varepsilon}
\def\phi{\varphi}
\def\la{\lambda}
\def\ka{\kappa}
\def\r{\rho}
\def\s{\sigma}
\def\z{\zeta}
\def\om{\omega}
\def\th{\theta}
\def\vth{\vartheta}
\def\vphi{\varphi}

 \def\be   {\begin{equation}}   
 \def\ee   {\end{equation}}
 \def\ba   {\begin{array}}     
 \def\ea   {\end{array}}
 \def\bea  {\begin{eqnarray}}   
 \def\eea  {\end{eqnarray}}
 \def\bean {\begin{eqnarray*}}  
 \def\eean {\end{eqnarray*}}

\def\A{{\cal A}}
\def\C{{\bf C}}
\def\D{{\cal D}}
\def\E{{\cal E}}
\def\F{{\cal F}}
\def\G{{\cal G}}
\def\H{{\bf H}}
\def\Hb{{\bf H}}
\def\h{{\cal H}}
\def\I{{\cal I}}
\def\J{{\bf J}}
\def\Jb{{\bf J}}
\def\K{{\cal K}}
\def\Kb{{\bf K}}
\def\L{{\cal L}}
\def\M{{\cal M}}
\def\N{{\cal N}}
\def\Mb{{\bf M}}
\def\Ob{{\bf O}}
\def\O{{\cal O}}
\def\P{{\cal P}}
\def\Q{{\cal Q}}
\def\R{{\bf R}}
\def\S{{\cal S}}
\def\T{{\rm T}}
\def\V{{\cal V}}
\def\W{{\cal W}}
\def\X{{\cal X}}
\def\Z{{\bf Z}}
\def\toro{{\bf T}}

\def\Ga{\Gamma}
\def\De{\Delta}
\def\La{\Lambda}
\def\Om{\Omega}
\def\Th{\Theta}

\def\pa{\partial}
\def\pd{\partial}
\def\d{{\rm d}}       %% derivative
\def\w{\wedge}
\def\xb{{\bf x}}
\def\x{\times}
\def\ox{\otimes}
\def\o+{\oplus}
\def\xd{{\dot x}}
\def\yd{{\dot y}}
\def\grad{\nabla}     %% gradient
\def\lapl{\triangle}  %% laplacian
\def\ss{\subset}
\def\sse{\subseteq}
\def\Ker{{\rm Ker}}
\def\Ran{{\rm Ran}}
\def\ker{{\rm Ker}}
\def\ran{{\rm Ran}}
\def\unm{{1 \over 2}}
\def\iff{{\rm iff\ }}
\def\all{\forall}
\def\LRA{\Leftrightarrow}
\def\<{\langle}
\def\>{\rangle}
\def\eor{{$\odot$}}
\def\EOR{~ \hfill {$\odot$}}
\def\shcomp{\odot}

\def\({\left(}
\def\){\right)}
\def\[{\left[}
\def\]{\right]}
\def\=#1{\bar #1}
\def\~#1{\widetilde #1}
\def\.#1{\dot #1}
\def\^#1{\widehat #1}
\def\"#1{\ddot #1}

\thispagestyle{empty}

\FirstPageHead{**}{**}{2005}{\pageref{gaeta-firstpage}--\pageref{gaeta-lastpage}}{Article}

\copyrightnote{2005}{G~Gaeta,~D~Levi,~R~Mancinelli}

\Name{Asymptotic symmetries of difference equations on a lattice}
\label{gaeta-firstpage}

\Author{Giuseppe GAETA}

\Address{Dipartimento di Matematica, Universit\`a di Milano, v. Saldini 50, I--20133 Milano (Italy); gaeta@mat.unimi.it}

\Author{Decio LEVI}

\Address{Dipartimento di Ingegneria Elettronica, Universit\`a di Roma Tre and INFN, Sezione di Roma III, via della Vasca Navale 84, I--00146 Roma (Italy); levi@fis.uniroma3.it}

\Author{Rosaria MANCINELLI}

\Address{Dipartimento di Fisica, Universit\`a di Roma Tre, Via della Vasca Navale 84, I--00146 Roma (Italy); mancinelli@fis.uniroma3.it}

\Date{Received: 20/12/2004}

\begin{abstract}
\noindent It is known that many equations of interest in Mathematical Physics display solutions which are only asymptotically invariant under transformations (e.g. scaling and/or translations) which are not symmetries of the considered equation. In this note we extend the approach to asymptotic symmetries for the analysis of  PDEs, to the case of difference equations.
\end{abstract}

\section{Introduction}

When we consider a  differential
equation  the analysis of symmetries leads, via a standard
procedure \cite{BlK,Gae,Olv,Ste,Win}, to the determination of
invariant solutions.

This method has been extended to consider {\it conditional
symmetries} \cite{ck,fush,lw,olver}, i.e. transformations which do not leave the equation
invariant, but such that some of its solutions are left invariant.
This notion can be further generalized to consider {\it partial
symmetries} where a subset of solutions, each of them not necessarily invariant, is mapped into itself by the transformation at hand  \cite{CGpar} (see also \cite{CiK} for a short review).

In a recent contribution \cite{GaM} the notion of {\it asymptotic
 invariant solutions} with respect to standard symmetries and 
conditional symmetries for differential equations has been introduced.  This theory
easily extends also to partial symmetries. The approach of
\cite{GaM} can be seen as a development of \cite{Gasy}, based in
turn on the renormalization group approach to differential
equations \cite{Bar1,Bar2,BK,CoE,Gol}. 
In this way we were able to explain the asymptotic scaling symmetry observed in numerical experiments analyzing
anomalous diffusion and reaction-diffusion equations \cite{MVV}. In this note, {\it asymptotic} is meant in the sense of 
``for large values of some dependent or independent variables''. 

Another interesting notion in this field is that of  {\it approximate symmetries} \cite{appr1,fush2}. Here one considers transformations which leave invariant the lower order approximations to the solution of a system depending on a small parameter. This notion may also mean an approximate symmetry for a series expansion near a fixed point \cite{CG,Gio}.

All these approaches and results have been introduced in the case of differential equations.
However, in many physical (or biological, chemical, etc.) models
-- and in numerical simulations of continuous phenomena -- one is
rather interested in difference equations on a lattice, and their
symmetry properties \cite{Dor,Flo,LTW,LVW,LeW2,Mae,Win2}. The
reader is referred to the bibliography contained in \cite{Win2} for
applications.

The purpose of the present note is to show that the approach of
\cite{GaM} also extends to equations defined on a lattice.

We will not try to be as general as possible, but rather consider, after a brief introduction to the generalities of equations on a lattice and their symmetries,
a well defined equation. We hope that, in this way, 
the general method will result clearer, and it should be easy for the
reader  to apply it to other equations of interest. In fact, we
will focus on reaction-diffusion equations \cite{CLV,CrH,Ebe}, and
in particular on the classical FKPP equation \cite{Fis,KPP,Mur}.

The only possible symmetries for discrete equations if we do not
want to change essentially the underlying lattice are discrete translations, 
rotations of the lattice variables and continuous transformations of 
the dependent variable.

We are
mainly interested here in equations whose solutions (as observed in
numerical experiments) are asymptotically well described by a {\it
travelling front}. We believe that the reader interested in more
complex asymptotic symmetries will easily understand how to deal
with them by comparing the present work and \cite{GaM}.

In Section \ref{de} we present the generalities on equations defined on a lattice while in Section \ref{s1} we will consider symmetries for equations defined on a lattice. 
In Section \ref{FKPP} we introduce the FKPP equation both in the continuous and on a lattice and analyze it so as to deduce the correct asymptotic behavior by requiring that asymptotically the system has the correct required symmetries. Section \ref{con} is devoted to a final discussion of the results and some concluding remarks.

\section{Difference equations on a lattice.} \label{de}

For simplicity we will consider a lattice $\La$ in $\R^2 = (x,t)$, and a difference equation for a real function $u(x,t)$ on it. Higher   dimensional lattices and matrix functions defined on them could be defined in the same way.  We will use the notation of
\cite{LeW2,Win2}.  Thus the points on the lattice, having
coordinates $(x_{m,n} , t_{m,n} )$, will be indexed by a couple of
numbers $(m,n) \in {\bf Z}^2$, and to each site is associated a
real variable $u_{m,n} \in \R$.

The lattice is  intrinsically  described by
assigning the difference between  neighboring points and will be of the form
\bea  \label{1.1}
x_{m+1,n} - x_{m,n} = \xi_{m,n} \ ,& \ x_{m,n+1} - x_{m,n} = \eta_{m,n}\, \\ \nonumber
t_{m+1,n} - t_{m,n} = \tau_{m,n} \ ,& \ t_{m,n+1} - t_{m,n} = \vartheta_{m,n} \ . 
\eea
These will be referred to as the {\it lattice equations}.
The set of values $\xi_{m,n}$ defines a function $\xi : \La \to \R$ via $\xi (x_{m,n} , t_{m,n},u_{m,n} ) = \xi_{m,n}$, and similarly for $\eta , \tau, \vth$.
By the notation $u_{m,n}$ we mean $u(x_{m,n},t_{m,n})$.
The shifts $(m,n) \to (m+1,n)$ and $(m,n) \to (m,n+1)$ correspond to $(x \to x + \xi (x,t,u) , t \to t + \tau (x,t,u)) $ and to $(x \to x + \eta (x,t,u) , t \to t + \vth (x,t,u) )$ respectively. The presence of $u$ allows for solution depending lattices.

We will then consider a difference equation for the variables $u_{m,n}$:
 \be \label{1.2}
 F_{m,n} [u]: \; \; \;  \mathcal{F}[x_{m,n}, t_{m,n}, \{ u_{m+j,n+i}
\}_{(i,j) \in \Z^2}]=0.
 \ee
Given some initial boundary conditions, this equation must allow us to get $u$ in all points of the lattice (past or future).

As an example of difference equation we can consider the  ``discrete heat equation'',
 obtained as the natural discretization (but only one of the possible ones) of $u_t = u_{xx}$,
\be \label{1.3}
 (\de t)^{-1} \ [ u_{m,n+1} - u_{m,n} ] \ = \ (\de x)^{-2} \ [ u_{m+1,n} - 2 u_{m,n} + u_{m-1,n} ] \ .
 \ee
where $\xi_{m,n}=\de x$, $\eta_{m,n}=0$, $\tau_{m,n}=0$ and $\vartheta_{m,n}=\de t$. 
We can express the discretizations in terms of an operator $\Delta_k$, a
\textit{difference operator of order $j-i$}. Acting with $\Delta_k$ on an
arbitrary smooth function $f_k=F(z_k)$ we have
\be \label{5.13}
\Delta_k f_k  =  \frac{1}{\delta_k} \sum_{\ell=i}^j a_\ell \, f_{\ell+k}\ , \
\sum_{\ell=i}^j a_\ell = 0 \ , \ \sum_{\ell=i}^j \ell \, a_\ell = 1 \ , 
\ee
where $\delta_k=z_{k+1}-z_k$.
In eq. (\ref{1.3}) the simplest discrete derivative on $t$, corresponding
to $j=1$ and $i=0$, has been considered.

\section{Symmetries} \label{s1}

A transformation $S : (x,t,u(x,t)) \to (x',t',u' (x',t'))$ will be a symmetry
of the lattice equation if it takes $\La \ss \R^2$ into itself.  It
will be a symmetry of the difference equation defined on $\La$ if
it is a symmetry of the lattice and leaves the 
equation $F_{m,n}[u] $  invariant. Most of the transformations for the lattice will be discrete symmetries.
If we  consider a uniform regular lattice $\La_R$, the point
$(m,n)$ will have coordinates $x_{mn} = x_{0,0}+m \de x$, $t_{mn} = t_{0,0}+n
\de t$ and  the lattice equations are
\bea \label{2.1}
x_{m+1,n} - x_{m,n} = \de x \ ,& \ x_{m,n+1} - x_{m,n} = 0 \, \\ \nonumber
t_{m+1,n} - t_{m,n} = 0 \ ,& \ t_{m,n+1} - t_{m,n} = \de t \ . 
\eea
We will denote by $p$ the ratio $p = \de x / \de t$ ($p\in$ {\bf R}).
On $\Lambda_R$ the discrete symmetries are easy to
determine.  They are
given by
$$ \cases{
\hat{T_x}: \ (x,t)  \to  (x+\de x , t ) & [shift in $x$,
equivalent to $(m,n) \to (m+1,n)$]; \cr \hat{T_t}: \ (x,t)   \to 
(x  , t + \de t ) & [shift in $t$, equivalent to $(m,n) \to
(m,n+1)$]; \cr \hat{B_x}: \ (x,t)  \to  (-x , t) & [inversion in
$x$, equivalent to $(m,n) \to (-m,n)$]; \cr \hat{B_t}: \ (x,t)  \to 
(x , - t) & [inversion in $t$, equivalent to $(m,n) \to (m,-n)$];
\cr \hat{R}: \ (x,t) \to (- p t , x/p ) & [rotation by $\pi/2$ with
a scale factor, \cr & \ equivalent to $(m,n) \to (-n , m)$].
\cr} $$ 
Moreover, if we accept $S \La_R \ss \La_R$ rather than $S \La_R= \La_R$, we  have the further trasformation
\be \label{2.2}
 {S}_q \ : \ (x,t) \to (q_1 x , q_2 t) \ , \ (q_1, q_2) \in {\Z} \ {\rm (discrete ~scaling)} \ . 
 \ee

Apart from the discrete symmetries given by the transformation written above we can introduce Lie symmetries which are better described in term of an  infinitesimal symmetry generator $\hat X$, which, taking into account that the lattice $\Lambda_R$ is subject just to discrete transformations, will involve only transformations of the dependent variable $u_{m,n}$
\be \label{2.3}
{\hat X}_{m,n} = Q(x_{m,n}, t_{m,n}, \{ u_{m+j,n+i} \}_{(i,j) \in \Z^2} )\partial_{u_{m,n}}.
\ee
Formula (\ref{2.3}) will be the infinitesimal generator of a symmetry for (\ref{1.2}) if 
\be \label{2.3a}
\sum_{j=-1}^1 \sum_{i=0}^1 {\hat X}_{m+j,n+i} {\mathcal F} |_{{\mathcal F} = 0} = 0
\ee
Eq. (\ref{2.3a}) is equivalent to the request that the flow generated by eq. (\ref{1.2}), when $u_{m,n} = u_{m,n}(\lambda)$, is compatible with 
\be \label{2.3b}
u_{m,n,\lambda} = Q(x_{m,n}, t_{m,n}, \{ u_{m+j,n+i} \}_{(i,j) \in \Z^2} )
\ee
Let us notice that the symmetries generated by the infinitesimal generator (\ref{2.3}), if $(i,j) \ne (0,0)$, are not Lie point symmetries but {\it generalized symmetries}.

Solutions invariant  with respect to the symmetries of infinitesimal generator (\ref{2.3}) are obtained by solving the difference equation (\ref{1.2})  together with the invariance condition $Q(x_{m,n},t_{m,n}, \{ u_{m+j,n+i} \}_{(i,j) \in \Z^2} ) = 0$. A particularly interesting class of function symmetries is when the function $Q$ takes the form
\bea \label{2.4}
Q &=& \phi(x_{m,n}, t_{m,n}, u_{m,n} ) - \xi(x_{m,n}, t_{m,n}, u_{m,n} ) \Delta_m u_{m,n} - \\ \nonumber
&&- \tau(x_{m,n}, t_{m,n}, u_{m,n} ) \Delta_n u_{m,n}
\eea
which in the continuous limit goes over to point symmetries.

When the transformation is not a symmetry, 
but there is an invariant solution,
we say that we have  a {\it conditional symmetry} for our problem.  
In this case to get the invariant solution
we can add to the equation the condition ${Q}=0$.

If we consider an equation $F_{m,n}[u]$ which does not explicitly depend on $x$ and $t$, it is invariant under $\hat{T}_x$ and $\hat{T}_t$ transformation. 
In this case we can consider  travelling wave solutions,
 depending on $\z=x-vt$. 
Defining $v=k (\de x / \de t)$ with $k \in \Z$, the travelling wave solutions can be 
obtained by imposing \cite{LVW} 
\be \label{2.4a}
{Q}= (\hat{T}_x^{-k} -\hat{T}_t)u_{m,n}. 
\ee
For $k=1$ this can be read, similarly as in the continuous case where
the condition is $ u_x + (1 / v) u_t = 0$,  
as ${Q}= \De_x^- u_{m,n}+ {1 \over v} \De_t^+ u_{m,n}$ 
 where $\De_x^-$ and $\De_t^+$ are the 'down' and  'up' discrete derivatives defined as
\be \label{2.5}
\De_t^+={T_t-1 \over \de t} \; \; \; \; \; \; \De_x^- ={1- T_x^{-1} \over \de x}.
\ee

As the regular lattice $\La_R$ is invariant under
$$ \hat{M} \ = \ \hat{T}_x^k \, \hat{T}_t: \  (m,n,u_{m,n}) \, \to \ (m+k,n+1,u_{m+k,n+1}) \ , $$ 
we can pass to the
 variables $(\z=x-vt, \tau=t, w(\z,\tau)=u(x-vt,t))$. Obviously ${Q}$ and $\hat{M}$ are related, as we have ${Q}=\hat{T_t}(\hat{M}^{-1}-1)u_{m,n}$.
So we can introduce two new indices $(\mu=m-kn, \nu=n)$.
 In these new independent variables the difference equation 
$F_{m,n}[u]$ will transform into  a new difference equation for the variable $w_{\mu,\nu}$, $\Phi_{\mu,\nu}[w]$.

The reduction of eq. (\ref{1.2}) by the symmetry (\ref{2.4a}) implies that $w_{\mu, \nu+1}=w_{\mu, \nu}$;
so, as $w_{\mu,\nu}$ is independent on $\nu$, we can write $w_{\mu,\nu} = A_\mu$
and reduce $\Phi_{\mu,\nu}[w]$ to $\tilde{\Phi}_\mu[A]$.
If this equation admits a
solution, it represents a travelling wave solution.

Let us now consider a continuous transformation on $A$ generated by the infinitesimal symmetry operator
\be \label{2.6}
\hat{X} = \Psi A_{\mu} \partial_{A_{\mu}},
\ee
where $\Psi$ is a constant. The infinitesimal transformation (\ref{2.6}) corresponds to the transformation ${\tilde A}_{\mu} = A_{\mu} + \lambda \Psi A_{\mu} + O(\lambda^2)$.
If the quantity
$$ W_{\mu} \ := 
\ \( \hat{X} \tilde{\Phi}_{\mu} \, [A] \) \ - \ \( \tilde{\Phi}_{\mu} \, [\hat{X} A ] \) $$
is not zero but goes to zero for large $\mu$, then $\hat{X}$ is an
asymptotic symmetry of $\tilde{\Phi}_\mu[A]$.

Finally, if there exist $\hat{X}$-invariant asymptotic 
travelling wave solutions of the original
equation $F_{m,n}[u]$,
we will say that $\hat{X}$ is an
\textit{asymptotic conditional symmetry} of $F_{m,n}[u]$.

\section{The discrete FKPP equation and its symmetries} \label{FKPP}

In the rest of this note we will focus on the
standard FKPP \cite{Fis,KPP,Mur} 
reaction-diffusion equation
\be \label{4.1}
 u_t \ = \ u_{xx} \ + \ u (1-u).
 \ee
See e.g. \cite{CLV,CrH,Ebe} for a discussion of its properties.
Here we are specially interested in its asymptotic solutions for
large $t$. For sufficiently localized non-negative initial data,
\be \label{t2.26} 
u(x,t=0)= {\Bigl \{} \begin{array}{ccc} \stackrel{x \to +\infty}{\rightarrow}  e^{- k_0 ( x + k_1)}, & ( k_0, k_1) >0, &  \mbox {for $x > 0$} \\ [3pt]
 = 1,&   & \mbox{for $x < 0$}. \end{array} 
\ee
the asymptotic solutions correspond to a stable front travelling with constant speed
and smoothly connecting the ``fresh'' (unstable) stationary state $u
= 0$ ahead of it and the ``exhaust'' (stable) stationary state $u =
1$ behind it.

The difference equation representing its simplest
discretization on a regular lattice $\La_R$ is given by
\bea \label{4.2}
F_{m,n}[u]: &&   {u_{m,n+1} - u_{m,n} \over \de t}  \ = \\ \nonumber
&& = \frac{ u_{m+1,n} - 2 u_{m,n} + u_{m-1,n}}{\de x^2 } 
 \ + \ u_{m,n} \, \( 1 - u_{m,n} \) \ .
\eea
Let us stress that $\de x$ and $\de t$ are small but finite parameters, representing the lattice spacing in the $x$- and in the $t$-direction respectively.

As noted above, in analyzing the symmetries of eq. (\ref{4.2}), we should
restrict our attention to transformations 
$S: (x,t,u (x,t) ) \to (x' , t' ,
u' (x',t') )$ which are also symmetries of the lattice equations (\ref{2.1}).

Let us first consider transformations which do not act on the $u_{m,n}$. It
is easy to see that shifts in $x$ and $t$ are symmetries of the
equation (\ref{4.2}). This is also the case for the inversion in $x$, but not for the inversion in $t$. Also, since $x$ and $t$
are intrinsically different here, the rotation $R$ is not a
symmetry of the equation.
We can recover the time inversion by considering the map $ B_t :
(x,t,u(x,t)) \to (x,-t,u(x,-t))$, corresponding to $(m,n,u_{m,n})
\to (m,-n,u_{m,-n})$. This represents a time inversion and
backward dynamics.

It is easy to see (by the same argument as for
the continuous equation) that there is no way to have a scaling
symmetry of the independent variables for eq. (\ref{4.2}).

\subsection{The discrete FKPP in a moving frame}

Following the results presented in Section \ref{s1}, a set of values $u_{m,n}$ will represent a travelling solution
with speed $v = k \de x / \de t = k p$ (with $k \in \Z$) if,
for any $j\in \Z$,
\be \label{4.3}
 u_{m+k j,n+j} \ = \ u_{m,n} \ . 
 \ee
The equivalent of passing to a moving frame of reference will be
passing to the independent variables
\be \label{4.4}
 \z = x - v t \ ; \ \tau = t.  
 \ee
The inverse of the change (\ref{4.4}) is given by  $t = \tau $, $x = \z + v \tau$.
It follows from
\bea  \nonumber
\z_{m,n} \ &=& \ x_{m,n} - v t_{m,n} \ = \ m (\de x) - { k \de x \over \de t} \, n \, \de t \ = \ (m - k n ) \de x \ , \\ \label{4.5}
\tau_{m,n} \ &=& \ t_{m,n} \ = \ n \, \de t \ .  
\eea
that the lattice equations, in terms of the new
coordinates,  are given by
\bea  \nonumber
\z_{m+1,n} - \z_{m,n} \ = \ \de x \ ,& \ \z_{m,n+1} - \z_{m,n} \ = \ - v \, \de t \ = \ - k \, \de x \ ; \\ \label{4.7}
\tau_{m+1,n} - \tau_{m,n} \ = \ 0 \ ,& \ \tau_{m,n+1} - \tau_{m,n} \ = \ \de t \ . 
\eea

The lattice point with coordinates $(\z = \mu \de x , \tau = \nu
\de t )$, has coordinates $(x = m \de x , t = n \de t)$ in the
original system, with $ \mu  = m - k n$, $\nu = n$.  The inverse
map is $m = \mu + k \nu$, $n = \nu$.

We will denote by $w_{\mu , \nu}$ the dependent variable associated to the
point of coordinates $\z = \mu \de x , \tau = \nu \de t$. We have
\be \label{4.8}
w_{\mu , \nu } \ \equiv \ u_{\mu + k \nu , \nu } ; \ u_{m,n} \ = \ w_{m - k n , n } . 
\ee

We can now use (\ref{4.8}) to express (\ref{4.2}) in the new variable
$w$. We get
\bea \label{4.9}
{w_{m-k(n+1),n+1} - w_{m-kn,n} \over \de t} & =& w_{m-kn,n} ( 1 - w_{m-kn,n} ) + \\ \nonumber 
&+&  {w_{m+1-kn,n} - 2
w_{m-kn,n} + w_{m-1-kn,n} \over \de x^{2} }.
\eea
Passing now to the indices $(\mu,\nu)$, we can rewrite  eq. (\ref{4.9}) as
\be \label{4.10}
\Phi_{\mu,\nu}[w]: \ {w_{\mu - k,\nu + 1} - w_{\mu,\nu} \over \de t } \ = \
{w_{\mu+1,\nu} - 2 w_{\mu,\nu} + w_{\mu-1,\nu} \over \de x^{2}
}  \ + \ w_{\mu,\nu} \, \( 1 - w_{\mu,\nu} \) . 
\ee
Eq. (\ref{4.10}) can be  simplified by recalling that $\de
t = {k \over v} \de x$. So the equation $\Phi_{\mu,\nu}[w]$ reads
\be \label{4.11}
\[ w_{\mu+1,\nu} - 2 w_{\mu,\nu} + w_{\mu-1,\nu} \] 
-{v \over k} \de x \[ w_{\mu - k,\nu+1} - w_{\mu,\nu} \]
+ \de x^2  w_{\mu,\nu} \, \( 1 - w_{\mu,\nu} \) \ . 
\ee
Let us recall that $\de x$ is a small but finite parameter,
representing the lattice spacing in the $x$-direction.

If we reduce eq. (\ref{4.11}) with respect to a proper combination of the translations, we get that the solution of (\ref{4.11}) will depend only on the $\z$ (not on the
$\tau$) coordinate, i.e.  
$w_{\mu,\nu+j} =w_{\mu,\nu}$ for all $j \in \Z$. 
If $w_{\mu,\nu}$ does not depend on the $\tau$ variable, 
we can as well
restrict to a fixed value of the $\nu$ index, and write simply
$A_\mu := w_{\mu,\nu}$.
Inserting this ansatz into
(\ref{4.11}) we get the {\it reduced difference equation}
\be \label{4.12}
 \tilde{\Phi}_\mu: \ \ \[ A_{\mu+1} - 2 A_{\mu} + A_{\mu-1} \] \ + \  \,{v \over k} \de x \[ A_{\mu} - A_{\mu-k}\] \ + \
\de x^2 \, A_{\mu} \, \( 1 - A_{\mu} \) \ = \ 0 \ . 
\ee

\subsection{Travelling fronts}

A solution to (\ref{4.12}) is, by construction, a travelling wave solution
to the difference FKPP equation (\ref{4.2}). In the continuum case, we
know this will actually be a travelling front. In the
moving frame coordinates $(\z,\tau)$, it will quickly tend to
zero for positive $\z$ and to one for negative $\z$. Moreover,
it is stable. If we start with a sufficiently regular initial
datum, the (non translation-invariant) solution will tend to this
moving front as time goes by.
Finally, the front solution is characterized by a sharp transition
between the two stationary states $u=0$ and $u=1$. The region with
small $u$ is described by an exponential, $u \simeq \exp{(- \a \z)}$ 
(with $1/\a$ a characteristic width).

The equivalent of the exponential behaviour in our discrete setting is
\be \label{5.1}
A_{\mu + 1} \ = \ (1- \a \de x) \, A_{\mu}.
\ee
In fact, as $\mu=\zeta / \de x$, 
$\lim_{\de x \to 0} A_{\mu}=A_0 \exp{(-\alpha \zeta)}$.
Solving eq. (\ref{5.1}) we get
 \be \label{5.2}
A_{\mu}=(1-\a \de x)^{\mu} A_0.
\ee

Putting formula (\ref{5.2}) in (\ref{4.12}), with $k=1$, we have
\be \label{5.3}
\de x^2 \left[{\a^2 - (v + \de x)\a + 1 - (1-\a \de x)^{\mu+1} A_0}\right]=0.
\ee
Eq. (\ref{5.3}) is not solvable for $\mu \to \infty$ if $\a<0$.
The case $\a>0$ gives us a 
monotonically decreasing solution for growing $\mu$.
For large $\mu$, we can leave out the last term of eq. (\ref{5.3})
and solve it for $\a$: 
\be \label{5.4}
\a^2 - (v + \de x)\a + 1 =0 .
\ee
This equation admits two real
solutions for $v + \de x >2$; 
studying the stability of the equation as in the continuous case,
one finds that the velocity is $v=2-\de x$. So the
front, observed in the numerical integration of the discrete FKPP on $\La_R$, 
would be just a little slower (of the order of $\de x$) than the front, which is 
solution of the FKPP equation. 

Note that we have just found an asymptotic symmetry.
In fact the choice of a scaling invariant solution (\ref{5.1}) is equivalent 
to set 
\be \label{5.5}
Q=  \a A_{\mu} + \De_{\z} A_{\mu} 
\ee
The corresponding transformations,  
 a scaling on $A$ and a translation on $\z$, are not a symmetry
 of the non-linear equation (\ref{4.12})
as $W_{\mu}=\hat{X} \^\Phi_\mu [A] - \^\Phi_\mu [\hat{X}A]=
- \a \de x^3 (1- \a \de x)A_{\mu+1}^2 \ne 0$.
 As for $\a<0$ $A_{\mu}$ is a decreasing function of $\mu$, 
we get that $\lim_{\mu \to \infty}{W_{\mu}}=0$.

\section{Discussion and Conclusions} \label{con}
In this note we have applied the method of asymptotic symmetries to a discrete version of the FPPK equation. In such a way we have shown that also in the discrete case the FKPP equation possesses an asymptotic scaling symmetry which gives the correct asymptotic behaviour of the system.

Long time ago Ibragimov and Fushchich \cite{appr1,fush2} introduced the notion  of approximate symmetry (for differential equations); this is based on some small parameter $\eps$ appearing in the equation, and is a symmetry "up to terms of order $\eps$". One could wonder what is the relation between approximate symmetries in this sense and our approach to the front solutions of the FKPP equation; note however that no explicit small parameter $\eps$ is present in the FKPP equation. A natural way to introduce such an $\eps$ would be to look for solutions whose amplitude is of order $\eps$, i.e. $u_{n,m} = \eps v_{n,m}$ (with $|v_{n,m}|$ of order one); proceeding in this way, however, the obtained scaling invariant solution will be small everywhere, not only asymptotically. Actually, in our approach the natural small parameter is inversely proportional to the group parameter generated by the asymptotic symmetry. 
Thus the small parameter comes into play only when we consider an additional object (the asymptotic symmetry). It appears, therefore, that our notion of asymptotic symmetry cannot be reconducted to the  notion of approximate symmetries of Ibragimov and Fushchich \cite{appr1,fush2}; further work would be needed for a better understanding of the relation between these two notions.

Work is in progress to obtain the asymptotic solutions for the anomalous reaction diffusion equation. This equation is nonautonomous in the independent variables, so that no point symmetry is available. In such a case the use of a renormalization of the lattice, provided by eq. (\ref{2.2}), might be important.
 
\section*{Acknowledgments}

RM was supported by INFM, DL  from  PRIN Project SINTESI-2002/2004 of the  Italian Minister for  Education and Scientific Research,  and GG and DL  from  the Project {\it Simmetria e riduzione di equazioni differenziali di interesse fisico-matematico} of GNFM--INdAM.


\begin{thebibliography}{39}

\parskip=0pt

\bibitem{appr1} Ba\u\i kov V A, Gazizov R K and Ibragimov N K,  Approximate symmetries, {\it (Russian) Mat. Sb. (N.S.)}, {\bf  136(178)} (1988) 435-450; translation in {\it Math. USSR-Sb.} {\bf  64} (1989) 427-441.

\bibitem{Bar1} Barenblatt G I, Scaling, self-similarity, and intermediate asymptotics, Cambridge University Press, Cambridge 1996.

\bibitem{Bar2} Barenblatt G I, Scaling, Cambridge University Press, Cambridge 2003.

\bibitem{BlK} Bluman G and Kumei S, Symmetries of differential equations, Springer, Berlin 1989.

\bibitem{BK} Bricmont J and Kupiainen A, Renormalization group and the Ginzburg-Landau equation, {\it Comm. Math. Phys.} {\bf 150} (1992), 193-208.

\bibitem{CLV} Cencini M, Lopez C and Vergni D, Reaction-diffusion systems: front propagation and spatial structures, in: The
Kolmogorov Legacy in Physics (LNP 636), Vulpiani A
and Livi R eds., Springer, Berlin 2003.

\bibitem{CiK} Cicogna G, Weak symmetries and adapted
variables for differential equations, {\it Int. J. Geom. Meth.
Mod. Phys.} {\bf 1} (2004), 23-31.

\bibitem{CG} Cicogna G and Gaeta G, Symmetry and perturbation theory in nonlinear dynamics (LNPm57), Springer, Berlin 1999.

\bibitem{CGpar} Cicogna G and Gaeta G, Partial Lie-point symmetries of differential equations, {\it J. Phys. A} {\bf 34} (2001), 491-512.

\bibitem{ck} Clarkson P A and Kruskal M D,  New similarity reductions of the Boussinesq equation, {\it J. Math. Phys.} {\bf 30} (1989), 2201-2213.

\bibitem{CoE} Collet P and Eckmann J P, Instabilities and fronts in extended systems, Princeton University Press, Princeton 1990.

\bibitem{CrH} Cross M C and Hohenberg P C, Pattern formation outside equilibrium, {\it Rev. Mod. Phys.} {\bf 65} (1993), 851-1112. 

\bibitem{Dor}  Dorodnitsyn V A, Transformation group in a space of difference variables, {\it J. Sov. Math.} {\bf 55} (1991), 1490-1517.

\bibitem{Ebe} Ebert U and van Sarloos W, Front propagation into unstable states: universal algebraic convergence towards uniformly translating pulled fronts, {\it Physica D} {\bf 146} (2000), 1-99.

\bibitem{Fis} Fisher R A, The Wave of Advance of Advantageous Genes, {\it Ann. Eugenics} {\bf 7} (1937), 355-369.

\bibitem{Flo} Floreanini R and Vinet L, Lie symmetries of finite difference equations, {\it J. Math. Phys.} {\bf 36} (1995), 7024-7042.  Floreanini R {\it et al.}, Symmetries of the heat equation on a lattice, {\it Lett. Math. Phys.} {\bf 36} (1996), 351-355.

\bibitem{fush} Fushchich V I, Serov N I and Chopik V I,  
Conditional invariance and nonlinear heat equations,  (Russian,  English summary) {\it Dokl. Akad. Nauk Ukrain. SSR Ser. A}  {\bf 86} (1988), 17-21.
Fushchich V I, Shtelen W M and Slavutsky S L, Reduction and exact solutions of the Navier-Stokes equations,  {\it  J. Phys. A}  {\bf 24} (1991), 971-984.

\bibitem{fush2} Fushchich W I and Shtelen W M,  On approximate symmetry and approximate solutions of the non-linear wave equation with a small parameter, {\it J. Phys. A} {\bf  22} (1989), L887-L890.

\bibitem{Gae} Gaeta G, Nonlinear symmetries and nonlinear equations, Kluwer, Dordrecht 1994.

\bibitem{Gasy} Gaeta G, Asymptotic symmetries and asymptotically symmetric solutions of partial differential equations, {\it J. Phys. A} {\bf 27} (1994), 437-451.

\bibitem{GaM} Gaeta G and Mancinelli R, Asymptotic symmetries of diffusion and reaction-diffusion equations, Preprint (2004).

\bibitem{Gio} Giorgilli A, Rigorous results on the power expansions for the integrals of a Hamiltonian system near an elliptic equilibrium point, {\it Ann. I.H.P. (Phys. Th\'eor.)} {\bf 48} (1988), 423-439.

\bibitem{Gol} Goldenfeld N, Martin O, Oono Y and Liu F, Anomalous diffusion and the renormalization group in a non-linear diffusion process, {\it Phys. Rev. Lett.} {\bf 65} (1990), 1361-1364.

\bibitem{GTW} Grundland A M, Tempesta P and Winternitz P,
Weak Transversality and Partially Invariant Solutions, {\it J.
Math. Phys.} {\bf 44} (2003), 2704-2722.

\bibitem{KPP} Kolmogorov A N, Petrovskii L G and Piskunov N S,
Etude de l'equation de la diffusion avec croissance de la
mati\'ere et son application \`a un probl\'eme biologique, {\it Bull. Moscow Univ. Math. Mech.} {\bf 1} (1937), 1-25.

\bibitem{LTW} Levi D, Tremblay S and Winternitz P, Lie point symmetries of difference equations and lattices, {\it J. Phys. A} {\bf 33} (2000), 8507-8524; 
Lie symmetries of multidimensional difference equations {\it J. Phys. A} {\bf 34} (2001), 9507-9524.

\bibitem{LVW} Levi D, Vinet L and Winternitz P, Lie group formalism for difference equations, {\it J. Phys. A} {\bf 30} (1997), 633-649.

\bibitem{lw} Levi D and Winternitz P, Non-classical symmetry reduction: example of the Boussinesq equation, {\it J. Phys. A} {\bf 22} (1989), 2915-2924.

\bibitem{LeW2} Levi D and Winternitz P, Continuous symmetries of discrete equations, {\it Phys. Lett. A} {\bf 152} (1991), 335-338; Symmetries and conditional symmetries of differential-difference equations, {\it J. Math. Phys.} {\bf 34} (1993), 3713-3730; Lie point symmetries and commuting flows for equations on lattices, {\it J. Phys. A} {\bf 35} (2002), 2249-2262.

\bibitem{Mae} Maeda S, The similarity method for difference equations, {\it IMA J. Appl. Math.} {\bf 38} (1987), 129-134.

\bibitem{Man} Mancinelli R, Asymptotic solutions of generalized diffusion and reaction-diffusion equations, unpublished.

\bibitem{MVV} Mancinelli R, Vergni D and Vulpiani A, Superfast front propagation in reactive systems with non-Gaussian diffusion, {\it Europhys. Lett.} {\bf 60} (2002), 532-538; 
Front propagation in reactive systems with anomalous diffusion, {\it Physica D} {\bf 185} (2003), 175-195.

\bibitem{Mur} Murray J D, {\it Mathematical Biology}, Springer, Berlin 1993.

\bibitem{Olv} Olver P J, Application of Lie groups to differential equations, Springer, Berlin 1986.

\bibitem{olver} Olver P J and Rosenau Ph,   Group invariant solutions of differential equations, {\it SIAM J. Appl. Math.} {\bf 47} (1987), 263-278.
  
\bibitem{Ste} Stephani H, Differential equations.  
Their solution using symmetries, Cambridge University Press 1989.

\bibitem{Win} Winternitz P, Lie groups and solutions of nonlinear PDEs, in: Integrable systems, quantum groups, and quantum field theory (NATO ASI 9009), Ibort L A and Rodriguez M A eds., Kluwer,
Dordrecht 1993.

\bibitem{Win2} Winternitz P, Symmetries of discrete systems,  in: Discrete Integrable Systems (LNP 644),  Grammaticos B, Kosmann--Schwarzbach Y and Tamizhmani T eds., Springer, Berlin, 2004.


\end{thebibliography}
\end{document}